\documentclass[useAMS,usegraphicx]{mn2e}
\usepackage{rotate}
\usepackage{rotating}
\usepackage{times}
\newif\ifAMStwofonts
\AMStwofontstrue

\def\gs{\mathrel{\hbox{\rlap{\hbox{\lower4pt\hbox{$\sim$}}}\hbox{$>$}}}}
\def\ls{\mathrel{\hbox{\rlap{\hbox{\lower4pt\hbox{$\sim$}}}\hbox{$<$}}}}

\def\xmm{{\it XMM-Newton}}

\def\et{{et al.\ }}

\def\mrk79{{Mrk~79}}
\def\pg08{{PG~0844+349}}

\def\3c{{3C~273}}
\def\rg{{\thinspace r_{\rm g}}}

\def\fexxv{{Fe~\textsc{xxv}}}
\def\fexxvi{{Fe~\textsc{xxvi}}}

\def\cm{{\rm\thinspace cm}}
\def\erg{{\rm\thinspace erg}}

\def\ga{{\rm\thinspace gauss}}

\def\keV{{\rm\thinspace keV}}

\def\s{{\rm\thinspace s}}
\def\ks{{\rm\thinspace ks}}

\def\ergpscmps{\hbox{$\erg\cm^{-2}\s^{-1}\,$}}

\def\ergcmps{\hbox{$\erg\cm\s^{-1}\,$}}

\title[How resonant absorption can mimic high-velocity outflows]
      {
How the effects of resonant absorption on black hole reflection spectra
can mimic high-velocity outflows
      }

\author[L. C. Gallo \& A. C. Fabian]
       {L. C. Gallo$^1$ 
and	A. C. Fabian$^2$ 
        \\ 
$^{1}$ Department of Astronomy and Physics, Saint Mary's University, 923 Robie Street, Halifax, NS, B3H 3C3, Canada \\
$^{2}$ Institute of Astronomy, University of Cambridge, Madingley Road, Cambridge CB3 0HA\\
}
\date{Accepted. Received. }
\pagerange{\pageref{firstpage}--\pageref{lastpage}}
\pubyear{2011}
\begin{document}
\maketitle
\label{firstpage}

\begin{abstract}
Narrow absorption lines seen in the $2-10\keV$ spectra of active galaxies and Galactic black holes are normally 
attributed to iron in high velocity outflows or inflows.  We consider the possibility that such features could arise
naturally in the accretion disc.  Resonant absorption by highly ionised iron (e.g. \fexxvi\ and \fexxv) in an
optically-thin plasma that is located above the disc and rotating with it could reproduce narrow features in
the reflection component of the spectrum as it emerges from the disc. 
Depending on the inclination of the disc and the exact geometry of the hot plasma
(e.g. does it blanket the disc or a ring) apparently narrow absorption features could be detected between $4-10\keV$. 
Such an explanation requires no high velocity outflow/inflow and is consistent
with a reflection-based interpretation for accreting black holes systems.

\end{abstract}

\begin{keywords}
accretion, accretion discs --
black hole physics --
relativistic processes --
line: formation, identification --
galaxies: active --
X-ray: galaxies 
\end{keywords}


\section{Introduction}
\label{sect:intro}

The high energy resolution and signal-to-noise provided by current X-ray telescopes has
led to reports of narrow absorption lines in the spectra of black hole systems.  These features,
assumed to arise from iron ($E = 6.4-6.97\keV$) can be highly redshifted (e.g. Turner \et 2002, 2004; Dadina \et 2005;
Nandra \et 1999, 2007; Longinotti \et 2007; Pounds \et 2003a, 2005) or blueshifted (e.g. Nandra \et 2007; Dadina \et 2005;
Pounds \et 2003b; Reeves \et 2009).  Many of these features are transient and their actual existence have been
called into question on statistical grounds (e.g. Vaughan \& Uttley 2008).  Others are persistent in that they have
been identified in subsequent observations or with other instruments.  
These features are normally unresolved with current CCDs, constraining their widths to be less than a few hundred
electon volts. 
The most common explanation is that these features
originate in high-velocity outflows and inflows (see Cappi 2006 for a review) in some cases reaching speeds in excess
of $0.1c$ (e.g. Reeves \et 2009; Tombesi \et 2010).

An alternative explanation for the redshifted features are resonant absorption lines due to highly ionised iron that
arise naturally in the accretion disc (Ruszkowski \& Fabian 2000).  A hot and diffuse plasma located above the accretion
disc will imprint resonant absorption lines on the reflection spectrum as it emerges from the disc.  The medium is 
rotating with the accretion disc thus experiences the same dynamical effects as material in the disc.  
Here we show that both redshifted and blueshifted features can occur and 
potentially shift features to apparently high velocities without the need to invoke high-velocity outflows.

This is a proof of concept work where we examine the possibility that resonant absorption by highly ionised iron could 
account for the narrow features over a wide range of energies by making use of the velocities already present in the disc.
Detailed computational modelling is left to future work.  
It is certainly plausible that narrow absorption features close to $7\keV$ are due to highly ionised outflows.  However,
we consider that alternative explanations for the wide range of velocities reported need to be investigated.
In the following section we present the general picture and motivation for the study.  In Section 3 we describe the
potential features that would appear depending on the geometry of the system.  We discuss our results in Section 4.

\section{The model and motivation}
\label{sect:picture}

\begin{figure*}
\begin{center}
\begin{minipage}{0.48\linewidth}
\scalebox{0.32}{\includegraphics[angle=0]{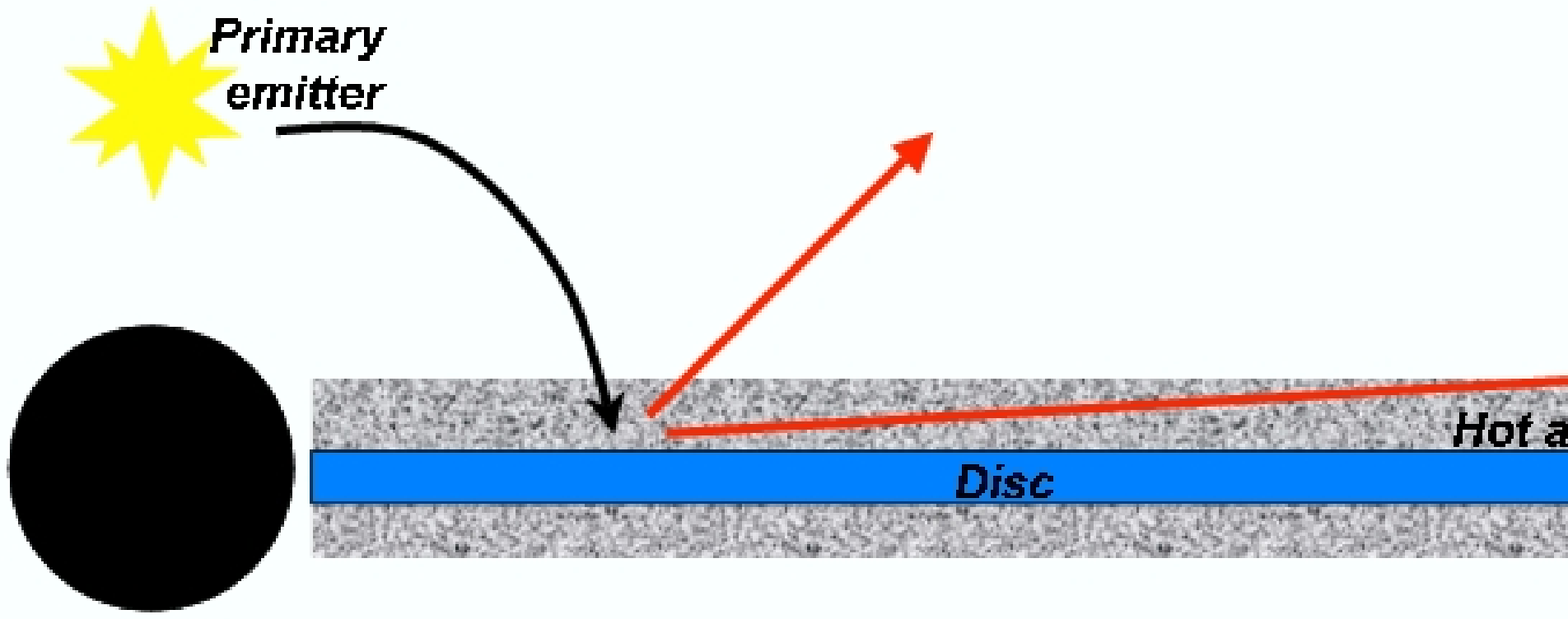}}
\end{minipage}  \hfill
\begin{minipage}{0.48\linewidth}
\scalebox{0.32}{\includegraphics[angle=270]{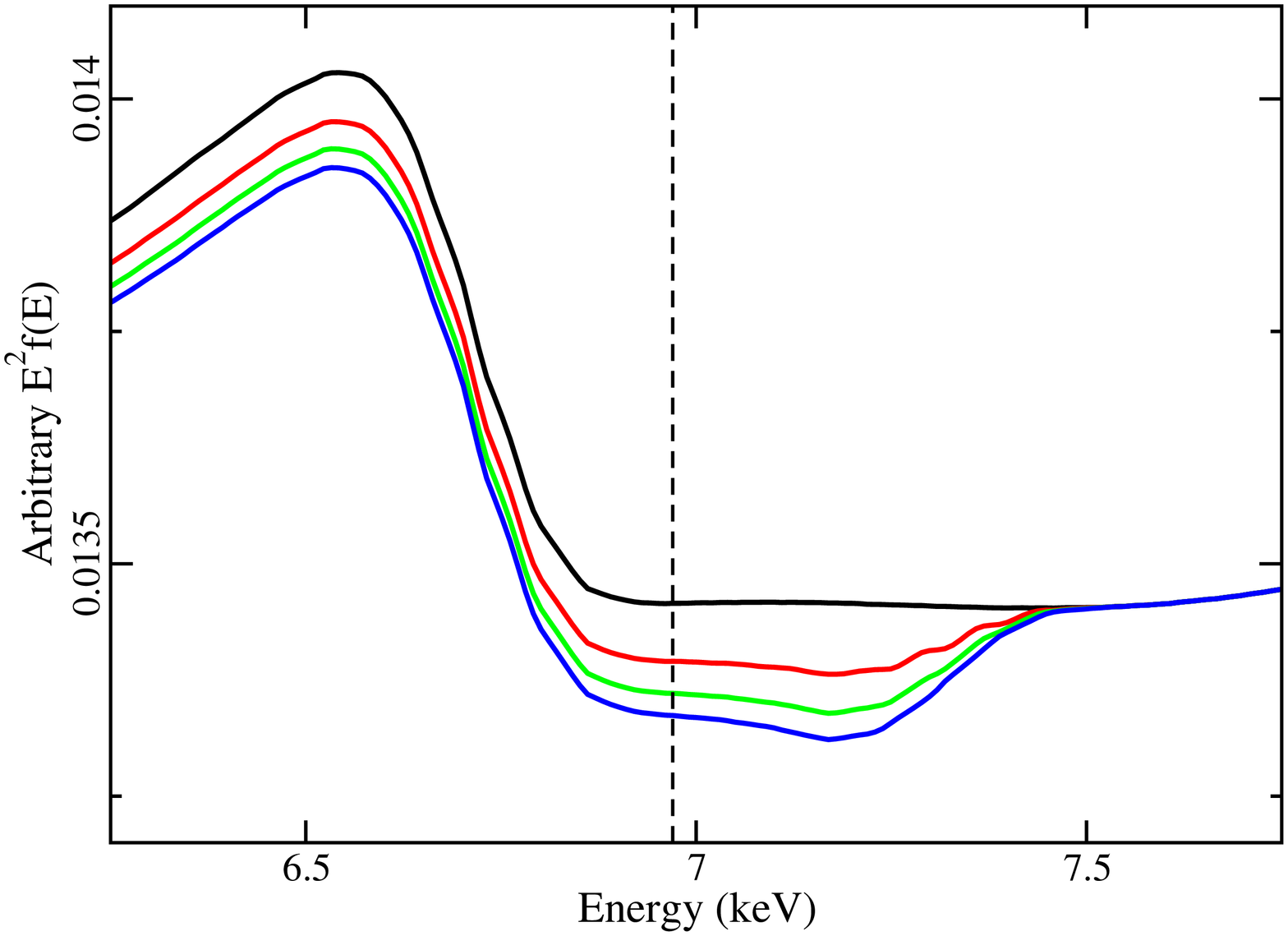}}
\end{minipage}
\end{center}
\caption{
Left:  The assumed geometry of the inner accretion disc.  The primary emitter is located on the
spin axis of the black hole and illuminates the standard accretion disc.  The reflection spectrum
(red curves) transverse the absorbing atmosphere above the disc.  Various paths through the hot atmosphere
can modify the depth of the absorbing features on the reflection spectrum.
Right: The modification to the blurred reflection spectrum due to \fexxvi\ absorption.
The black curve is the unabsorbed spectrum.  The red, green and blue curves correspond
to optical depths of 0.1, 1, and 10, respectively.
}
\label{fig:res}
\end{figure*}
\begin{figure*}
\begin{center}
\begin{minipage}{0.48\linewidth}
\scalebox{0.32}{\includegraphics[angle=270]{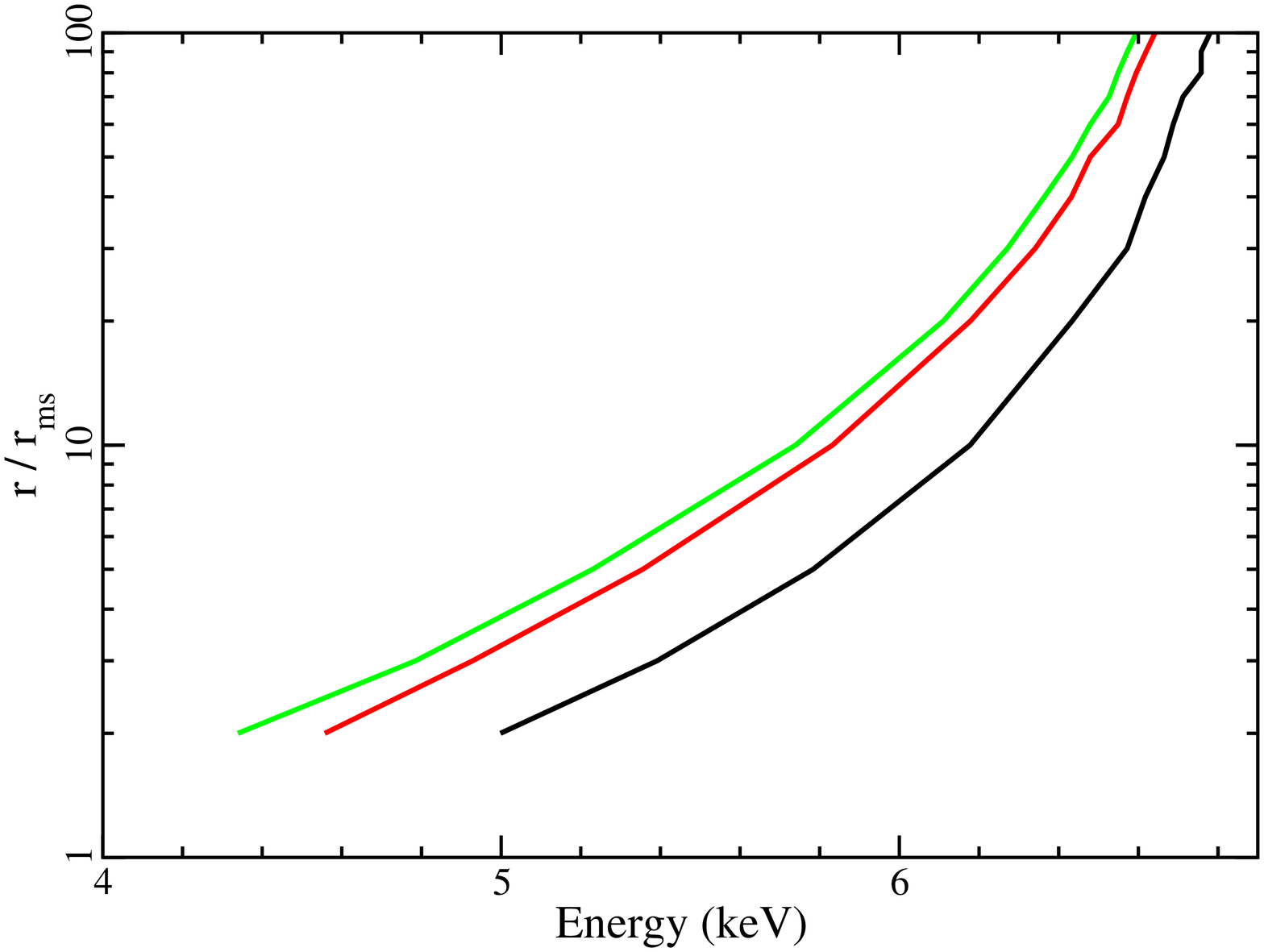}}
\end{minipage}  \hfill
\begin{minipage}{0.48\linewidth}
\scalebox{0.32}{\includegraphics[angle=270]{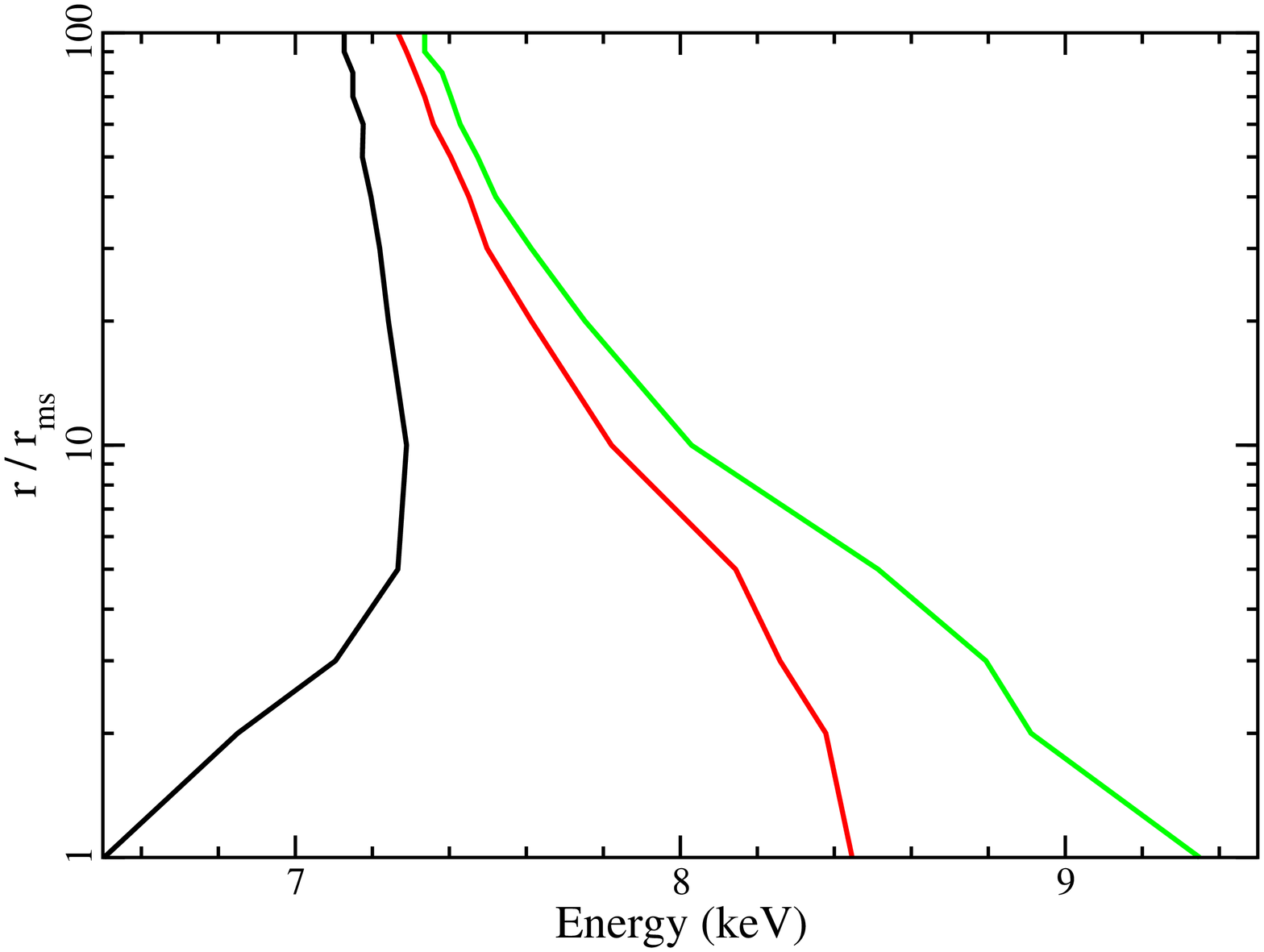}}
\end{minipage}
\end{center}
\caption{
Left:  The peak energy of the red wing of a narrow $6.97\keV$ line blurred by motions in the accretion
disc as a function of distance at which the absorption originates from a black hole with a spin parameter of $a=0.7$.  The distance is given relative
to the radius of marginal stability ($r_{ms}$).  The black, red and green line correspond to disc inclinations
of 30, 60, and 85 degrees, respectively.  The measurements are cut-off below $2r_{ms}$ as it is not possible to
isolate the peak of the red wing.
Right: Same as the left panel, but for the peak of the blue wing.
}
\label{fig:lines}
\end{figure*}

We adopt the standard accretion disc picture that has developed over recent years.  
The optically thick and geometrically thin
accretion disc is illuminated by a primary power law emitter that is located on the spin axis of the black hole
(Fig.~\ref{fig:res} left panel).  
We note that the location of the primary emitter is arbitrary and does not need to be on the spin axis of the
black hole, however from recent X-ray studies (e.g. Wilkins \& Fabian 2011) and gravitational microlensing observations of AGNs 
(e.g. Chartas \et 2009) 
we know the primary emitter must be compact and centrally concentrated.

Located above the accretion disc is an optically thin, highly ionised plasma that is corotating with the disc.
The reflection spectrum (i.e. reflection continuum and 
fluorescent lines, Ross \& Fabian 2005) emitted from the disc will cross through the plasma and may be subject to resonant 
absorption as it emerges from the disc.  Since the absorbing material is corotating with the disc the resonant absorption
features are subject to the
same kinematic and gravitational effects influencing the reflection spectrum.

Various lines-of-sight could influence the significance of the resonant features as the optical depth
through the absorbing medium changes (Fig.~\ref{fig:res} left panel).  This will alter the depth of the
feature as is shown in Fig.~\ref{fig:res} (right panel). 

As a simple test to examine if such a model could reproduce narrow absorption features in the appropriate
energy band we considered the observed energies of the red and blue peaks of an intrinsically narrow absorption
profile (Gaussian) at the rest energy of H-like iron (\fexxvi\ Ly$\alpha$, $6.97\keV$).  The line is broadened by Doppler and
relativistic effects using the {\tt kerrconv} model in {\tt XSPEC} (Brenneman \& Reynolds 2006) for a black hole 
spin parameter of $a=0.7$ (Fig.~\ref{fig:lines}).  Depending on the inclination of the disc, features attributed
to narrow \fexxvi\ absorption could appear anywhere from $\sim 4\keV$ to over $9\keV$.  The energy range could expand
to even lower energies if one considers absorption by \fexxv\ ($E=6.7\keV$) or higher energies if one considers
\fexxvi\ Ly$\beta$ ($E=8.25\keV$).

\section{Simulations}

In this section we consider various geometries of the model in Sect.~\ref{sect:picture} and examine
the appearance of the absorption imprinted on the reflection spectrum.  The direct component of the primary
emitter viewed by the observer is not absorbed by the plasma.  We assume the
hot atmosphere is highly ionised and consider resonant absorption only from hydrogen-like (\fexxvi\ Ly$\alpha$) and
helium-like (\fexxv) iron.  
The scattering is treated as simple line absorption, that is once the photon is scattered out of the line of sight 
it is neglected.
When considering two lines, for simplicity we treat both features as
equally strong.

We adopt a modest AGN spectrum for our trials.  The power law photon index is $\Gamma=2$.  The
disc is ionised with $\xi=100\ergcmps$ and solar iron abundances (Morrison \& McCammon 1983).  
The inner and outer disc edges are set
at $r_{in}=1.25\rg$ ($=1.25$~GMc$^{-2}$) and $r_{out}=100\rg$, respectively, and the emissivity index for the 
disc is $q=3$.
For each case considered, we simulate the appearance the spectrum could have when viewed during a $100\ks$
exposure with the \xmm\ pn detector.
The $2-10\keV$ flux is  $\sim10^{-11}\ergpscmps$.  The simulated spectra and models are shown in the AGN
rest-frame.

\subsection{Fully covered disc -- blanket absorption}
\label{sect:blanket}

\begin{figure*}
\begin{center}
\begin{minipage}{0.48\linewidth}
\scalebox{0.32}{\includegraphics[angle=270]{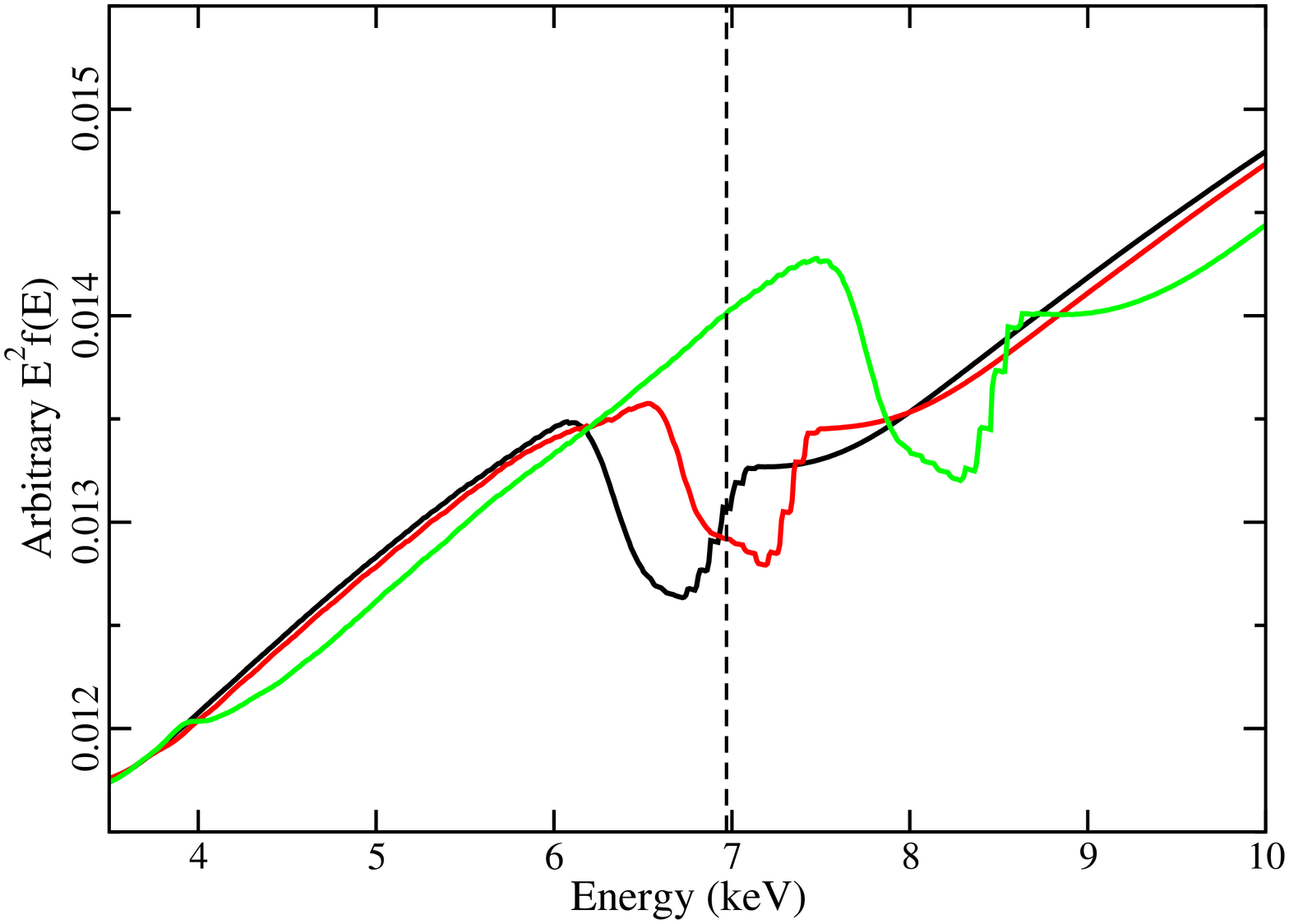}}
\end{minipage}  \hfill
\begin{minipage}{0.48\linewidth}
\scalebox{0.32}{\includegraphics[angle=270]{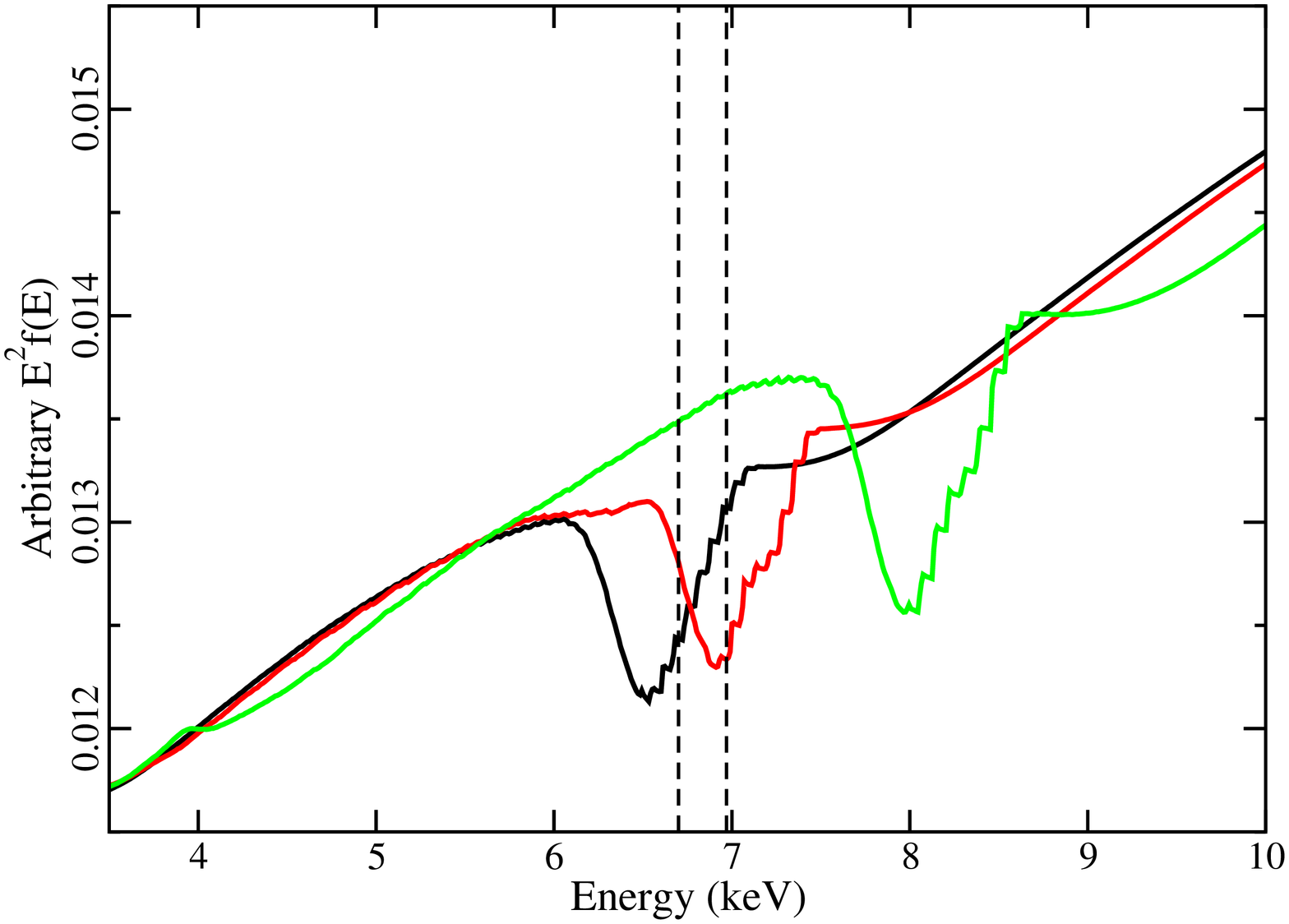}}
\end{minipage}
\end{center}
\caption{Left panel: The modification of the reflection spectrum due to \fexxvi\ absorption
that covers the entire disc.
The black, red, and green curves correspond to inclination angles of 
0, 30, and 60 degrees, respectively.  
Right panel: Same as the left panel, but with the absorption originating from \fexxvi\ and \fexxv.
The vertical dashed lines mark the position of
\fexxv\ ($6.7\keV$) and \fexxvi\ ($6.97\keV$).
}
\label{fig:blanket}
\end{figure*}
\begin{figure}
\rotatebox{270}
{\scalebox{0.32}{\includegraphics{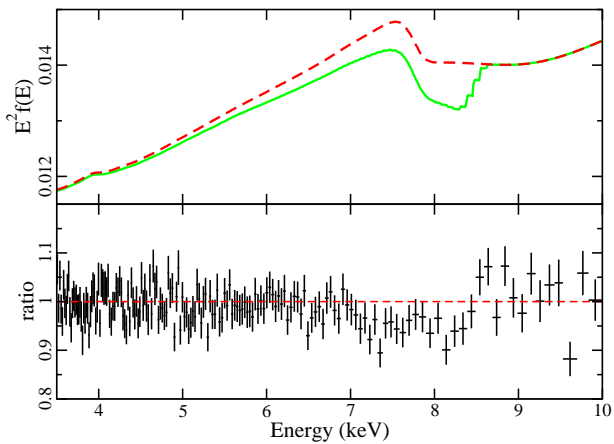}}}
\caption{Upper panel:  The green curve is the model corresponding to \fexxvi\ absorption that covers a
disc with an inclination of 60 degrees (same as green curve in Fig~\ref{fig:blanket}).  The red-dashed curve is the
absorption removed model.  
Lower panel:  The ratio between the two curves in the upper panel based on a $100\ks$ \xmm\ simulation. 
}
\label{fig:simblanket}
\end{figure}

The first case examined builds on the situation in Fig.~\ref{fig:res}.  
We consider an accretion disc that
is blanketed by the ionised plasma.  The plasma has a line optical depth of $\tau=1$ 
at all radii
and is irradiated with the same
emissivity profile as the disc.  
Fig.~\ref{fig:blanket} depicts the spectrum for a plasma made up of 
H-like iron (left panel) and both H- and He-like iron (right panel) resonant absorption at different 
line-of-sight inclinations. 

As seen in the left panel of  Fig.~\ref{fig:blanket} such features can be shifted and appear significantly broad and deep as they
blend with the iron absorption edge at 7.1 keV.  In the case of \fexxvi\ absorption viewed
at 60 degrees inclination, the feature can be redshifted well over $8\keV$ in the rest frame.
In Fig.~\ref{fig:simblanket} we demonstrate how such a feature may appear in an \xmm\ pn observation after
one has presumable modelled the reflection spectrum correctly.

If absorption were attributed to \fexxvi\ and \fexxv, the situation would resemble the right panel of 
 Fig.~\ref{fig:blanket}.  The lines would be blended due to relativistic broadening of each and would not
be distinguished as two features.
 
\subsection{Absorption in an annulus -- variable inclination }
\label{sect:ivar}

\begin{figure*}
\begin{center}
\begin{minipage}{0.48\linewidth}
\scalebox{0.32}{\includegraphics[angle=270]{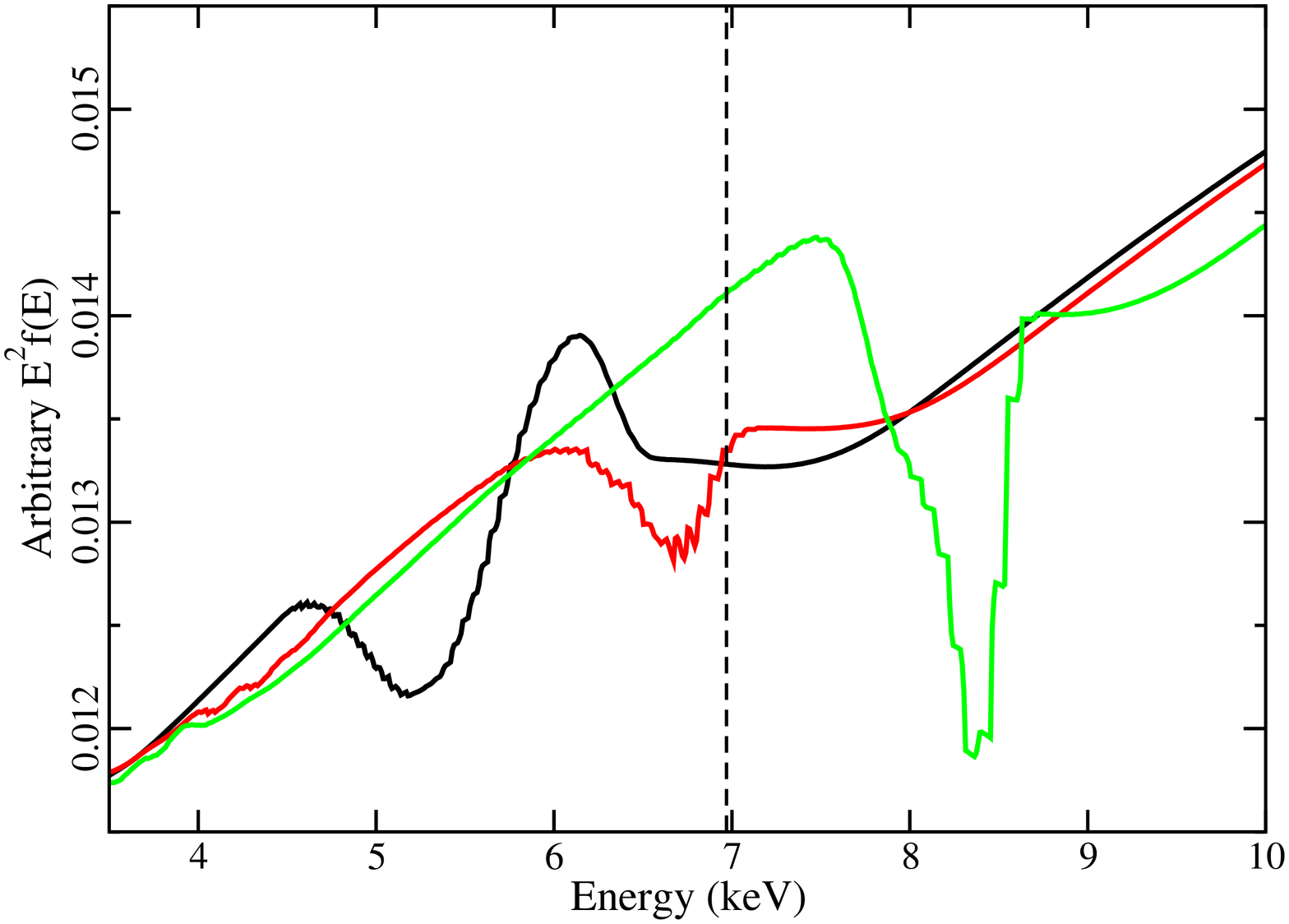}}
\end{minipage}  \hfill
\begin{minipage}{0.48\linewidth}
\scalebox{0.32}{\includegraphics[angle=270]{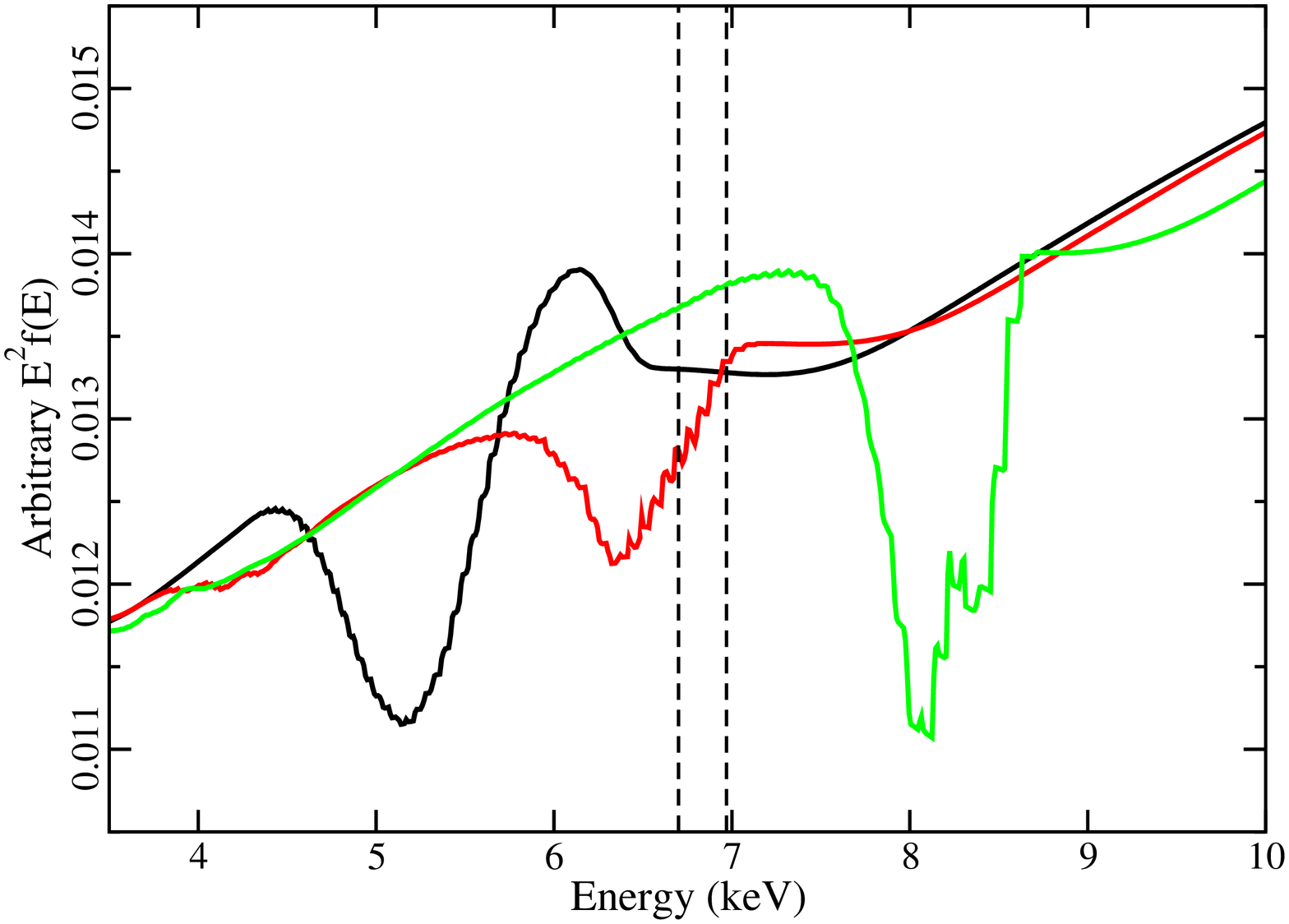}}
\end{minipage}
\end{center}
\caption{Left panel: The modification of the reflection spectrum due to \fexxvi\ absorption
that originates in a ring of material between $5-7\rg$..
The black, red, and green curves correspond to inclination angles of 
0, 30, and 60 degrees, respectively.  
Right panel: Same as the left panel, but with the absorption originating from \fexxvi\ and \fexxv.
The vertical dashed lines mark the position of
\fexxv\ ($6.7\keV$) and \fexxvi\ ($6.97\keV$).
}
\label{fig:ringI}
\end{figure*}

\begin{figure*}
\begin{center}
\begin{minipage}{0.48\linewidth}
\scalebox{0.32}{\includegraphics[angle=270]{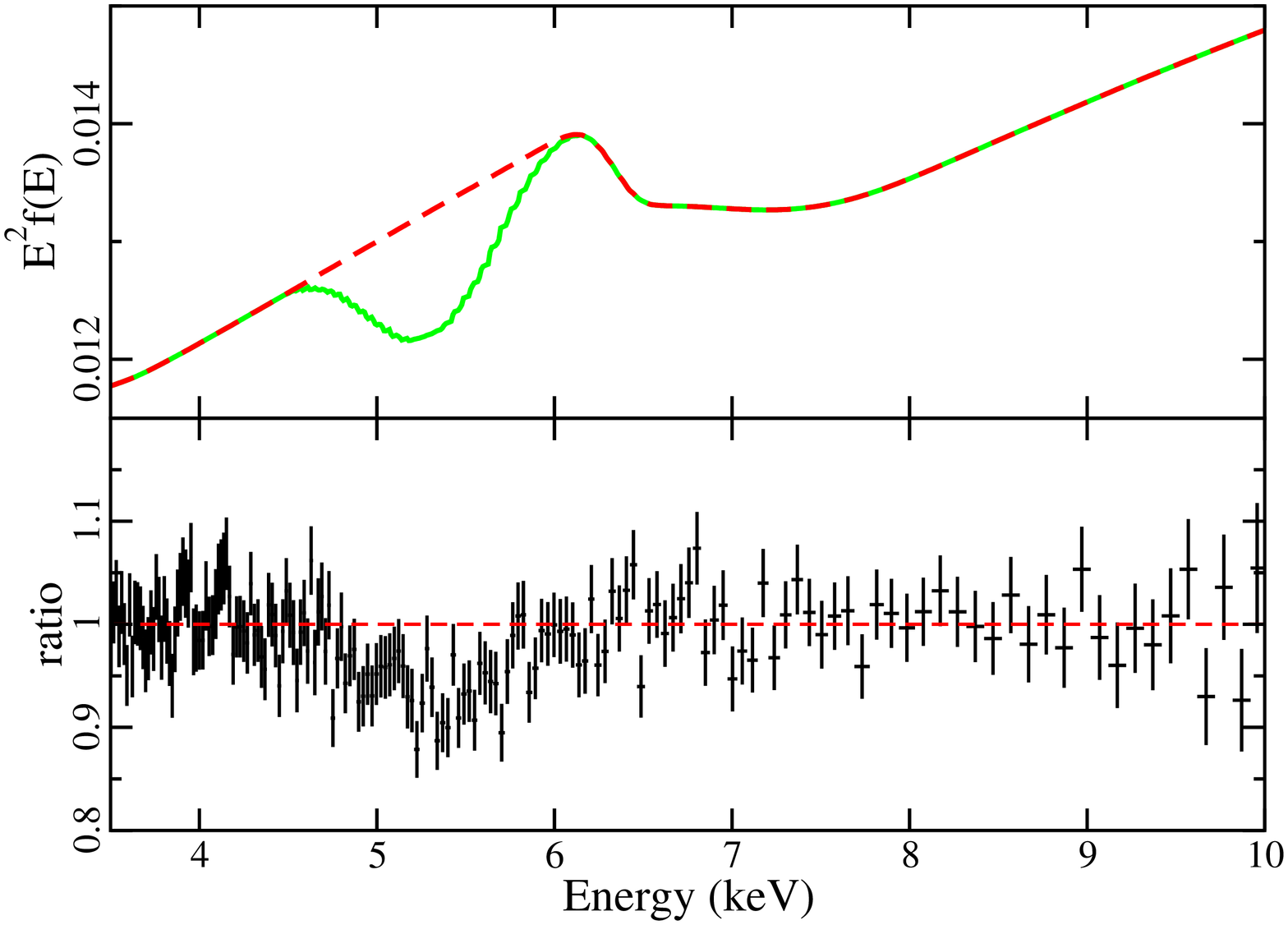}}
\end{minipage}  \hfill
\begin{minipage}{0.48\linewidth}
\scalebox{0.32}{\includegraphics[angle=270]{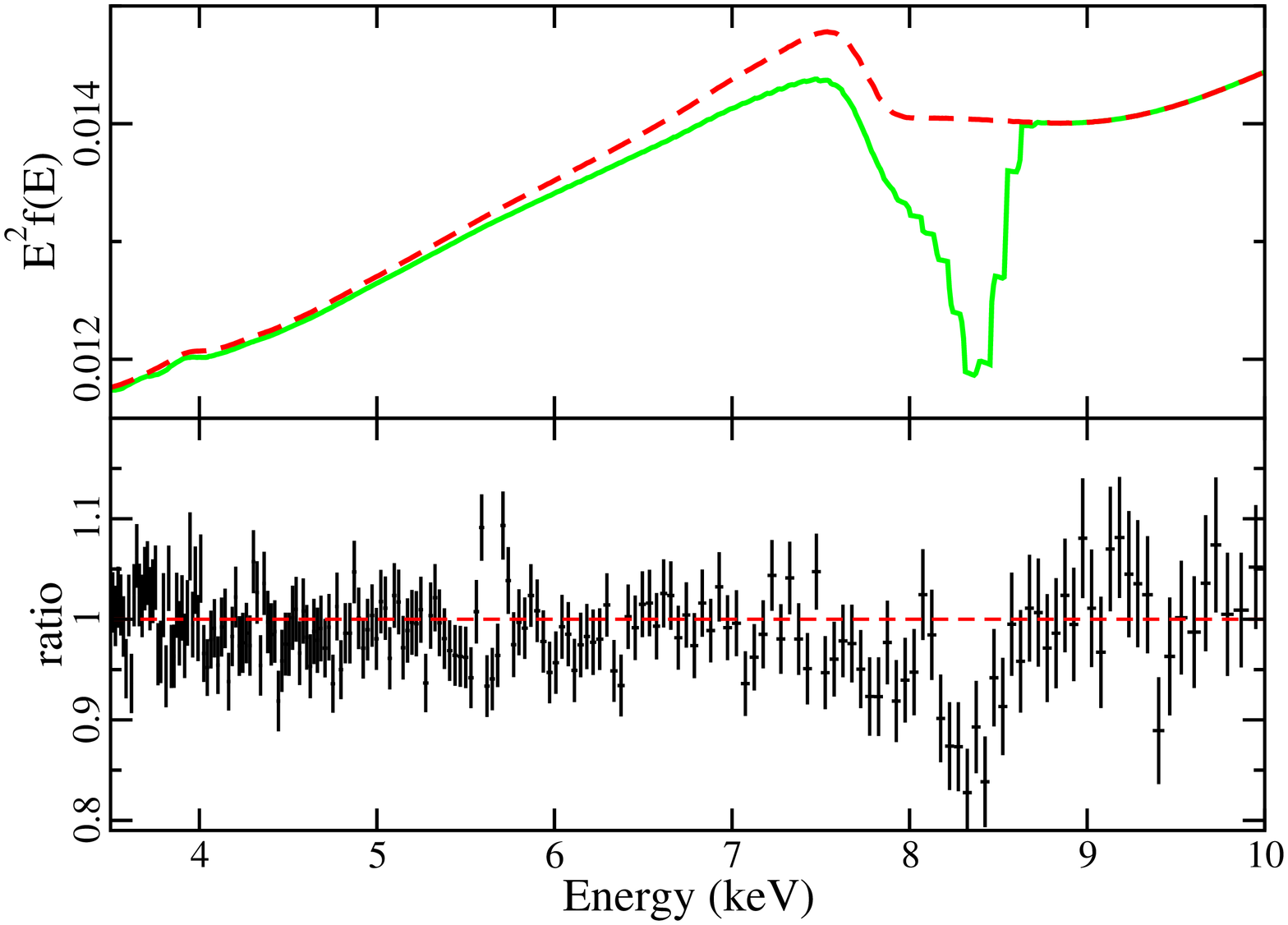}}
\end{minipage}
\end{center}
\caption{Left panel:  In the top panel, the green curve is the model corresponding 
to \fexxvi\ absorption from a ring between $5-7\rg$.
The disc has an inclination of 0 degrees (same as the black curve in Fig.~\ref{fig:ringI} left). 
The red-dashed curve is the
absorption removed model.  
In the lower panel, the ratio between the two curves based on a $100\ks$ \xmm\ simulation is shown. 
Right panel:  Same as left panel, but for an disc inclination of 60 degrees.
}
\label{fig:simringI}
\end{figure*}

In this and Section~\ref{sect:xvar} we still adopt the lamppost model described above, but consider 
the situation in which the absorption is enhanced in
a certain region above the disc effectively forming an annulus of absorbing material.  
This could originate
from a hot spot above the disc illuminating a certain region, or anisotropies in disc leading to increased
concentration of hot plasma, or enhanced structure in the disc that amplifies the reflection continuum.  
In the simulations shown in Fig.~\ref{fig:ringI}, the absorption is
localised between $5-7\rg$ from the black hole and viewed at different inclinations.  Depending on the
inclination, significant absorption lines can be seen between $\sim 5-8.5\keV$ in the rest-frame.
Detectable features are predicted in the \xmm\ pn simulation (Fig.~\ref{fig:simringI}). 

\subsection{Absorption in an annulus -- variable distance }
\label{sect:xvar}
\begin{figure*}
\begin{center}
\begin{minipage}{0.48\linewidth}
\scalebox{0.32}{\includegraphics[angle=270]{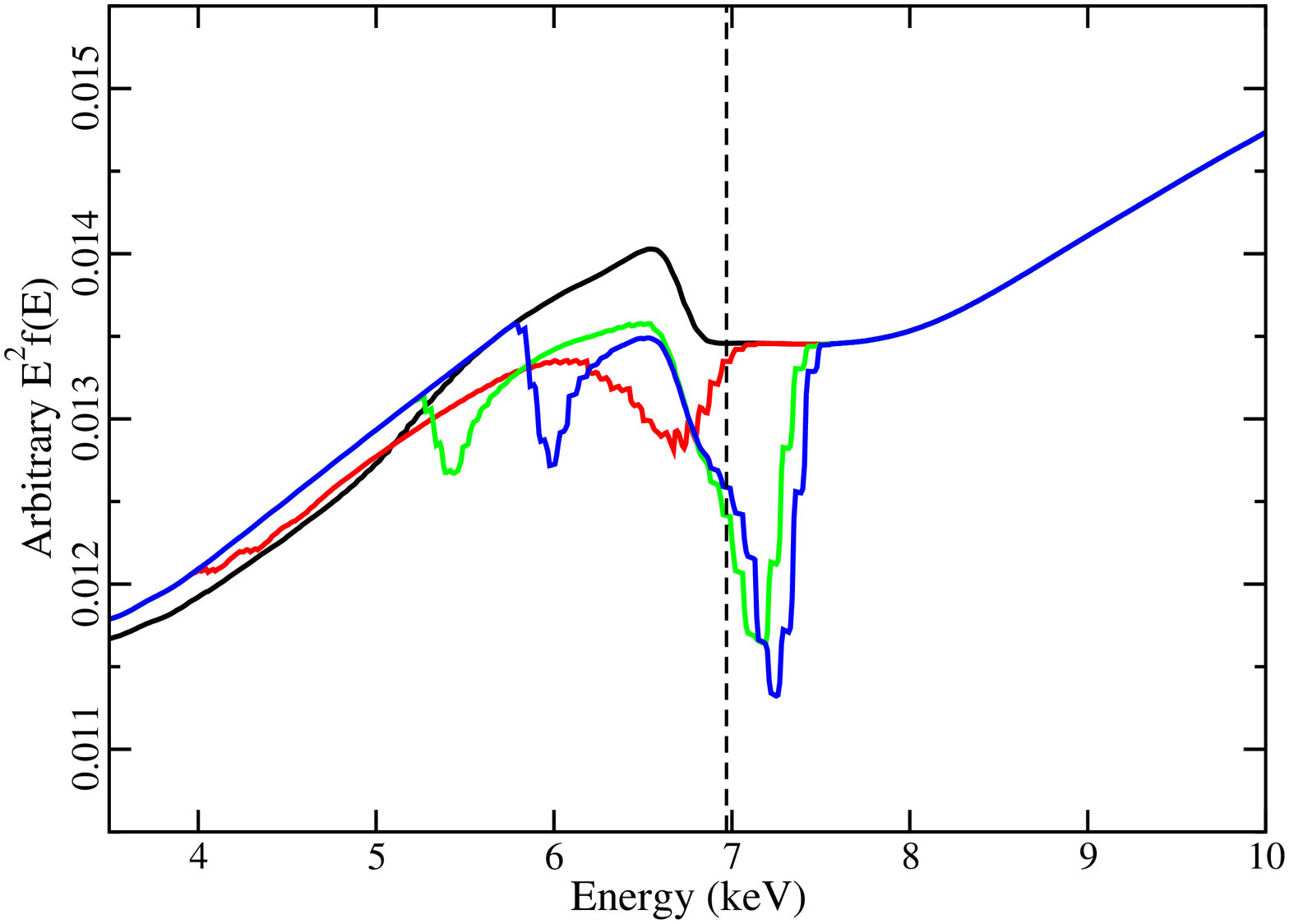}}
\end{minipage}  \hfill
\begin{minipage}{0.48\linewidth}
\scalebox{0.32}{\includegraphics[angle=270]{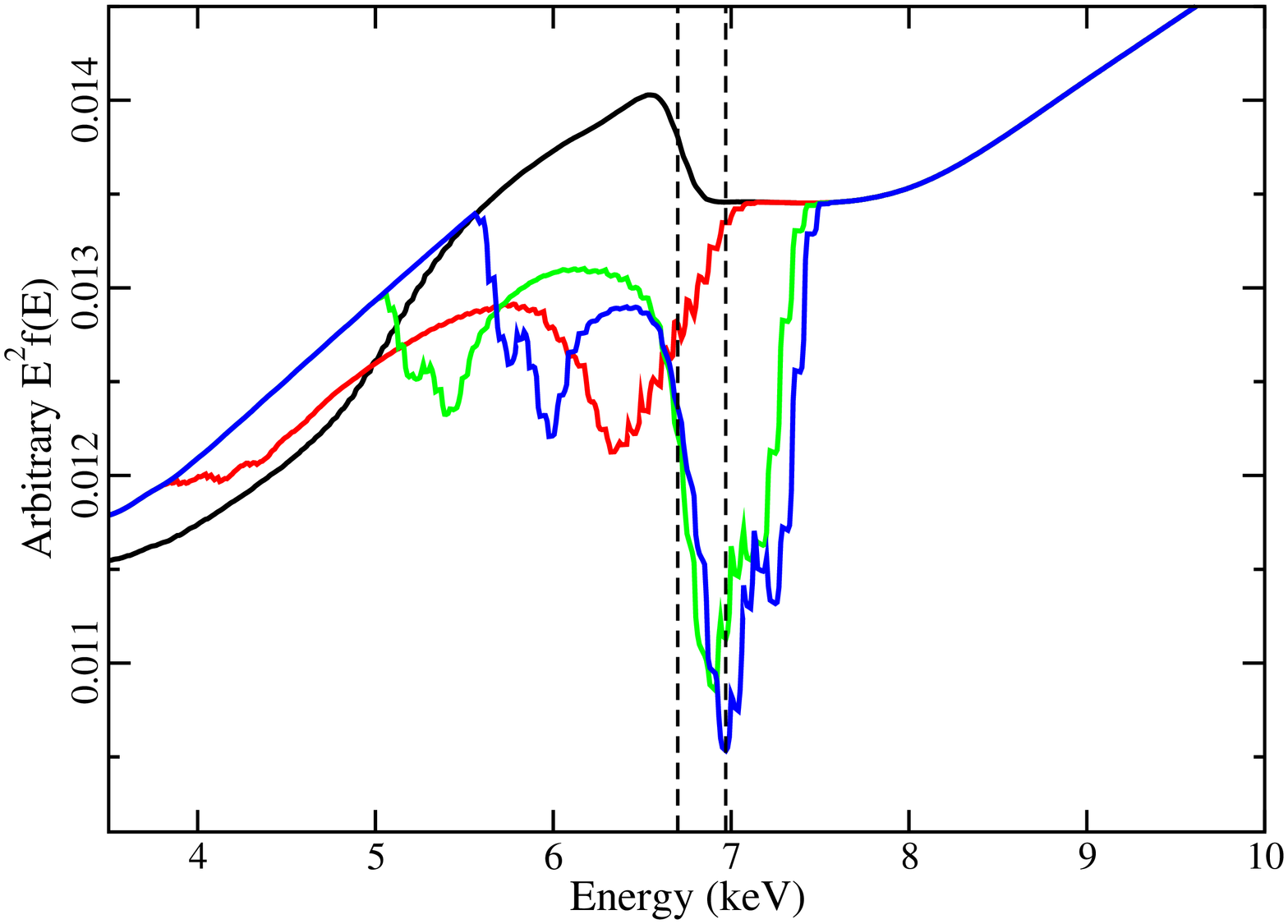}}
\end{minipage}
\end{center}
\caption{Left panel:  The modification of the reflection spectrum due to \fexxvi\
absorption in a ring of plasma arising at various distances along the disc.
The black, red, green, and blue curves correspond to rings at $1.25-3$, $5-7$,
$12-13$, and $20-21\rg$, respectively.  The disc is inclined at 30 degrees.
Right panel: Same as left panel, but with absorption arising from \fexxvi\ and \fexxv.
}
\label{fig:ringD}
\end{figure*}

\begin{figure}
\rotatebox{270}
{\scalebox{0.32}{\includegraphics{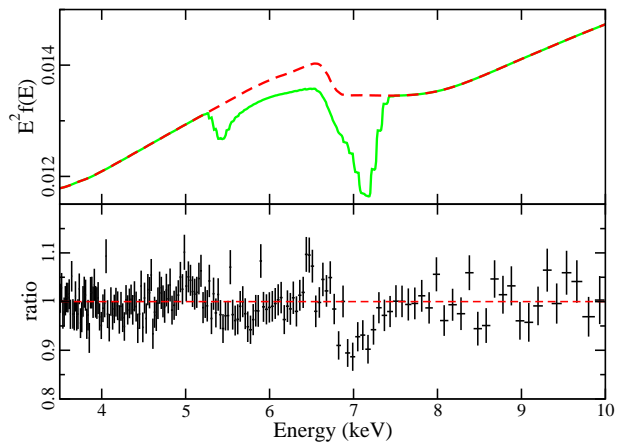}}}
\caption{Upper panel:  The green curve is the model corresponding to \fexxvi\ absorption arising in a 
ring between $12-13\rg$ and a disc inclination of 30 degrees inclination (green curve in Fig.~\ref{fig:ringD}).
The red-dashed curve is the
absorption removed model.  
Lower panel:  The ratio between the two curves in the upper panel based on a $100\ks$ \xmm\ simulation. 
}
\label{fig:simringD}
\end{figure}

The final situation examined corresponds to an absorbing annulus at various distances from the
black hole.  At distances less than $\sim5-7\rg$ the absorption profile is rather blurred (see also  
Section~\ref{sect:ivar}),
but as the distance increases and general relativistic effects are less dominant, Doppler effects prevail and the line 
profiles begins to take on the inverted disc line appearance.
In Fig.~\ref{fig:ringD}, we examine a disc inclined 30 degrees with absorption coming from different distances.
The double-peaked profile becomes apparent at distances beyond $\sim12\rg$.  The blue edge of the profile
is driven by the inclination of the disc and is shifted to higher energies as the disc inclination
increases (see Fig.~\ref{fig:lines}).  
The simulation in Fig.~\ref{fig:simringD} depicts \fexxvi\ absorption from a ring between $12-13\rg$.  The single
feature could easily mimic multiple absorption features.

\section{Discussion and conclusions } 
\label{sect:dis}

We examine the possibility that narrow absorption features regularly observed in the spectra
of AGN and attributed to fast, ionised inflows and outflows, could arise naturally from resonant absorption
of the reflection spectrum making use of velocities already present in the disc.  
We consider various geometries (e.g. rings and blankets) for the hot plasma
that is corotating with the accretion disc and subject to the same kinematic and relativistic effects as
the reflection spectrum.  In all cases, we demonstrate that absorption features could be easily 
detected between $\sim4-9\keV$ in typical \xmm\ observations.  
We note that the observed energy range could
be significantly expanded by adding the contribution of other transitions (e.g. \fexxvi\ Ly$\beta$ at $E=8.25\keV$).

Notable from the simulations in Section 3 is that the absorption features are not genuinely narrow, but
have significant width.  However, the CCD resolution provide by current instruments limits our ability
to distinguish features narrower than a few hundred eV.  Calorimeter observations with {\it Astro-H} and
{\it Athena} will likely be capable of resolving such features and thereby distinguishing models.

The work here is highly simplified and we only consider resonant absorption by \fexxvi\ and \fexxv.
More complicated situations, in which elemental abundances are non-solar (or variable)
and/or there exists greater diversity in composition of the plasma are easy to envision.  
Line of sight effects are also treated simply here.  If there is a composition gradient,
in the atmosphere then long sight lines through the hot plasma (e.g. low inclinations) can produces 
very complicated spectra with many features (e.g. the spectrum of the x-ray binary GRO J1655--40, Miller \et 2008).
Long sight lines will also mean that different species experience different blurring effects and a 
specific species will be associated with a specific distance from the black hole.
The purpose of this work was to demonstrate a concept leaving more complex situations for further study.


\section*{Acknowledgments}

LCG thanks colleagues at IoA, Cambridge for their hospitality.
ACF thanks the Royal Society for support. 
We thank the referee for providing comments that helped to clarify the paper.




\bsp
\label{lastpage}
\end{document}